\documentclass{article}

\usepackage{PRIMEarxiv}
\usepackage{graphics}
\usepackage[usenames,dvipsnames]{color}
\usepackage{subfigure}
\usepackage{epsfig} 
\usepackage{listings}
\usepackage{color}
\usepackage[framemethod=tikz]{mdframed}
\usepackage{enumitem}
\usepackage{comment}

\usepackage{amssymb}
\usepackage{amsmath}
\usepackage{pifont}
\title{Phases, morphologies, and transitions in a membrane model for
       the endoplasmic reticulum}
\author{Jaya Kumar Alageshan$^1$, 
         Yashodhan Hatwalne$^2$, Rahul Pandit$^1$ \\
        $^1$ Centre for Condensed Matter Theory, Department of Physics, \\ 
           Indian Institute of Science, Bangalore 560012, India. \\
        $^2$ Raman Research Institute, C.V. Raman Avenue, \\
           Sadashivanagar Bengaluru 560080, India
        }
\begin{document}
\maketitle

\begin{abstract}
We introduce a novel model, comprising self-avoiding surfaces and incorporating edges
and tubules, that is designed to characterize the structural morphologies and transitions
observed within the endoplasmic reticulum (ER). By employing discretized models, we
model smooth membranes with triangulated surfaces, and we utilize numerical variational
methods to minimize energies associated with periodic morphologies. Our study obtains
phases, their morphologies, and their transitions and examines their dependence on
the membrane chemical potential, the line tensions, and the positive Gaussian curvature stiffness. 
By starting with diverse topological structures, we explore shape variations by using Surface Evolver, while maintaining fixed topology. Notably, we identify the region of parameter space where the model displays lamellae, with a lattice of helical edges connecting the layers; this resembles structures that have been observed in the rough ER.
Furthermore, our investigation reveals an intricate phase diagram with
periodic structures, including flat lamellar sheets, sponge phases, and
configurations comprising tubules with junctions, which are akin to the morphology of the
smooth ER. An estimation of lattice parameters is achieved through fluctuation analysis.
Significantly, our model predicts a transition between homotopically equivalent
lamellae, with helical edges and configurations featuring tubules with junctions.

\end{abstract}

\section{Introduction}

\begin{figure}[!ht]
    \centering
    \includegraphics[scale=0.15]{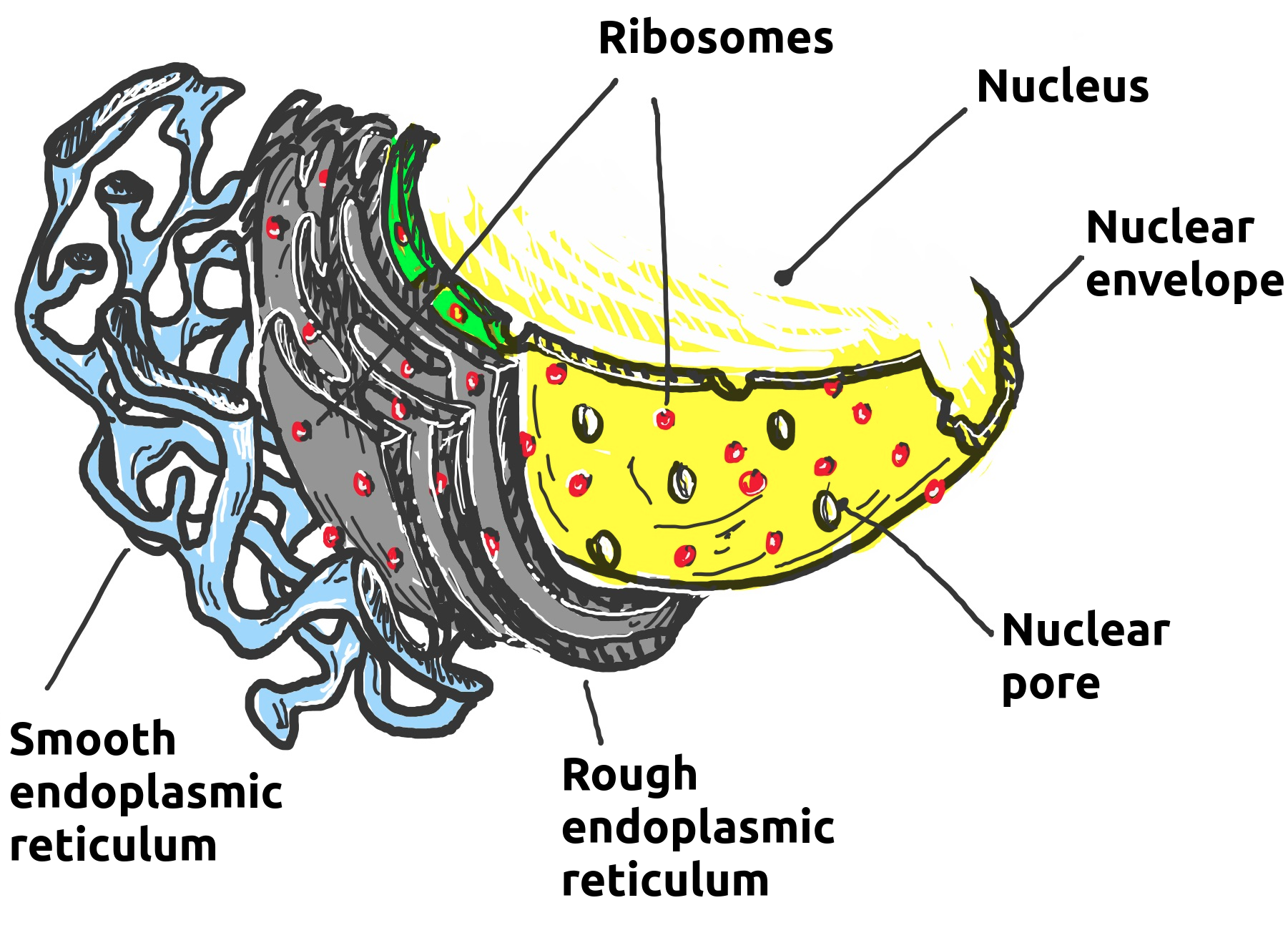}
    \caption{A schematic diagram illustrating the smooth and rough endoplasmic reticula (SER and RER), ribosomes,
    the cell nucleus, nuclear pores, and the nuclear envelope.}
    \label{Fig:ER}
\end{figure}

The endoplasmic reticulum (ER) surrounds the nucleus in most eukaryotic cells
and plays a crucial role in cellular structure and function. Its intricate
structure and diverse functions make it a cornerstone of cellular activity.
The basic building blocks of the ER are a network of membranous tubules and flattened
sheets called cisternae, which occur throughout the cytoplasm of eukaryotic
cells~\cite{Review} and exhibit distinct curvature, lipid composition, and protein
organization~\cite{Cell}. The interconnected tubular structures extend from the cisternae
(see Fig.~\ref{Fig:ER}). The tubules are dynamic and flexible structures that branch out
and fuse to form the complex three-dimensional network of the ER. The biophysics of the ER
encompasses the study of its structure, physical properties, and the molecular
dynamics underlying its functions. In the ER, the following two main regions have
been identified:

\begin{itemize}

    \item The {\it Rough Endoplasmic Reticulum} (RER) often appears as flattened
    cisternae and has a rough appearance under an electron microscope because of
    ribosomes on its outer surface~\cite{Ribosome}. These ribosomes are responsible
    for protein synthesis and the RER helps in the folding and also the
    modification of proteins within its lumen~\cite{ProteinFold}. Furthermore,
    the newly synthesized proteins' structural integrity and correct folding are
    ensured by chaperone proteins within the ER~\cite{Chaperone}. The lumen of the
    RER is contiguously connected to the lumen of the
    nuclues~\cite{Review}. 

    \item Away from the nucleus, the ER breaks down into a filamentous structure
    that forms the {\it Smooth Endoplasmic Reticulum} (SER). The SER lacks
    ribosomes on its surface, so it has a smooth appearance~\cite{Ribosome} and
    consists predominantly of tubules. It is involved in various metabolic
    processes, such as lipid synthesis~\cite{Lipid} and detoxification of
    toxins~\cite{Detox}. In particular, in muscle cells, the SER stores and
    releases calcium ions that are necessary for muscle contraction~\cite{Muscle2}.
    Calcium ions also play a pivotal role in cell signalling. therefore, the ER 
    plays a crucial role in their regulation and the proper functioning of the cell~\cite{Calcium}. The SER
    produces phospholipids and steroids that are essential for cell-membrane structure
    and hormone production~\cite{Steroid}.
    
\end{itemize}

The ER is dynamic and continually undergoes remodelling via the fusion and the
fission of tubules and cisternae; thus, it adapts to changes in cellular physiology and
structure. Proteins such as reticulons and REEPs are responsible for ER shaping
and dynamics~\cite{ER_Protein}, and they maintain the ER morphology and influence
its biophysical properties. The protein insertion, trafficking, and signalling
processes within the lipid bilayer of the ER membranes influence its mechanical
properties, fluidity, and curvature~\cite{Curvature}. In turn, these biophysical
properties influence the movement of vesicles and molecules within the ER.

The ER membrane is subject to active fusion and fission processes~\cite{Chen,Borgese}. 
There is a traffic of protein-carrying vesicles in the
ER-Golgi-plasma membrane, through the secretory pathways~\cite{MCB}. Typically,
about $1000$ vesicles leave the ER every second~\cite{Wei}. Despite this strong
remodelling activity, which is energetically driven by the consumption of ATP/GTP,
the ER has a well-defined stable lamellar structure~\cite{Schwarz}. The fusion
and fission processes occur predominantly at the outer interface between the
smooth ER and the cell plasma~\cite{Review}; so we hypothesize that local
equilibrium is valid in the bulk of the ER. Our findings demonstrate that this
hypothesis leads naturally to the structures of the ER that have been 
uncovered in recent experiments~\cite{Terasaki}.

Freeze-fracture studies~\cite{Terasaki} have established that the RER 
comprises stacks of flat sheets connected by screw dislocations with helical
edges (see Fig.~\ref{Fig:Helicoid}(a)). To avoid geometric frustration, the
constituents of the RER must have an equal number of left- and right-handed
dislocations; a two-dimensional (2d) section, normal to the axes of the
dislocation yields a square lattice with alternating handedness, as
shown in Fig.~\ref{Fig:Helicoid}(b). A morphological transition, from the flat
lamellar sheets to lamellae with helical edges occur as the lattice parameter,
i.e., the distance between the helical edges goes to infinity. We note, in passing,
that this transition is akin to the TGB-SmA (Twist Grain Boundary-Smectic-A) transitions in lyotropic
liquid crystals~\cite{K1}.

The ER is an essential and versatile cellular structure that plays a pivotal
role in various cellular functions. Thus, understanding the biophysics of the ER
is of paramount importance as it governs the organelle's function and interactions with
other cellular components. It is important to delve into the physical properties of ER
membranes, protein dynamics, and molecular processes, which occur within this
organelle; this is essential (a) for gaining insights into cellular physiology
and disease mechanisms, particularly those that are related to ER stress disorders, and
(b) for the development of targeted therapies. The ER works in conjunction with
other organelles, such as the Golgi apparatus, vesicles, and mitochondria,
to orchestrates a complex array of molecular activities,
significantly contributing to the overall functionality and health of
eukaryotic cells. 

In this study, we concentrate on the structural aspects of the ER, and organise the remaining part of this article as follows: In Sec.~\ref{sec:Model} we present our model.  Then, in Sec.~\ref{Sec:SE}, we give the details of the numerical shape variation that we employ to find the minimal-energy configuration for each of the ER
morphologies by using Surface Evolver~\cite{SE}. Finally, in Sec.~\ref{sec:Results}, we present the phase diagrams and morphologies we obtain. The concluding Sec.~\ref{sec:Conclusion} is devoted to  a discussion of the significance 
of our results and possible limitations of our study.

\section{The Model}
\label{sec:Model}


The ER has a complex multi-scale structure, with diverse morphologies and
functions that span length scales ranging from nanometers to micrometres.
The ER membranes are composed of lipid bilayers that have a thickness
$\simeq 4-6$ nanometers. Protein folding, chaperone interactions,
and post-translational modifications and other molecular processes occur
within the ER lumen and determine the double-bilayer thickness, which is typically
$\simeq 30-50$ nanometers~\cite{DB_Size}. Furthermore, the ribosomes
that decorate the outside of the RER are $\simeq 20-30$ nanometers in
diameter; they are responsible for protein synthesis, which occurs at the nanometer
scale within the ER cisternae. Throughout the cytoplasm, the SER forms an extensive network, which 
extends over micrometre scales in most eukaryotic cells. Vesicles within the ER network are involved in the 
transport of proteins, lipids, and other molecules, within the ER and between the ER and other
cellular compartments, such as the Golgi apparatus. The sizes of these
vesicles can vary, but their diameters lie typically in the range $25-50$
nanometers. 

The ER's diverse length scales, which are vital for its functions, allow it to perform tasks
ranging from molecular-level protein synthesis and modification to coordinating cellular
processes over larger distances. This ability to cover multiple scales enables efficient
communication and coordination within the cell and underscores the ER's importance in 
maintaining cellular homeostasis. The ER is complex: it has many types of
protein molecules, with non-trivial properties, embedded within its membrane.

\begin{figure}[!ht]
    \centering{
	\includegraphics[scale=0.17]{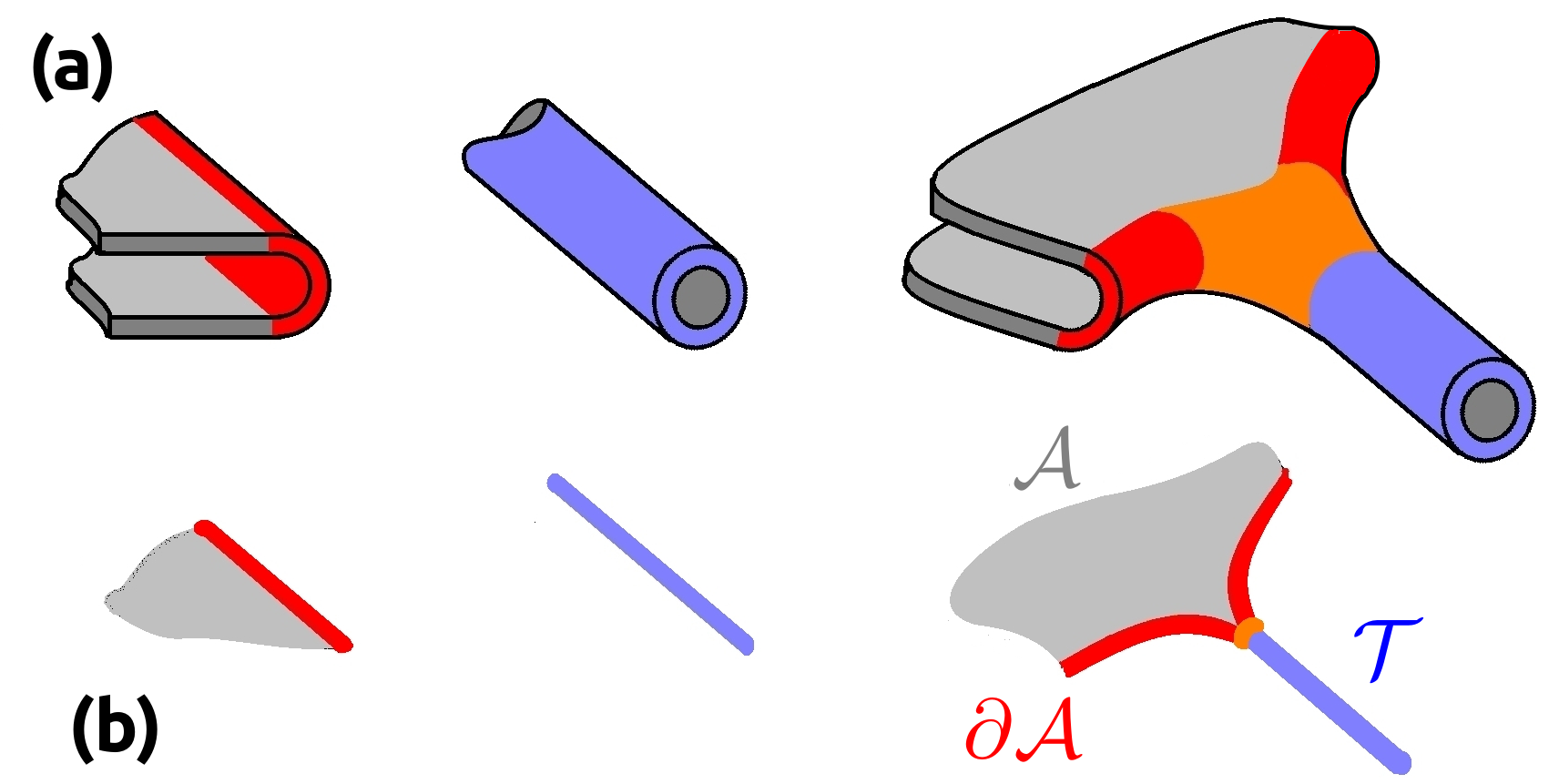}}
	\caption{(a) A colour-coded schematic diagram of the elements of the double bilayer membrane in our model:
	   {\color{gray}($\mathcal{A}$)}- 2D surface bulk; 
          {\color{red}($\partial \mathcal{A}$)}- surface edge; 
	   {\color{blue}($\mathcal{T}$)}- tubule;
          {\color{orange}(orange)}- junction of 
	   tubule and edge; (b) the corresponding equivalent 2d surface and 
	   1d curve in our model.} 
	\label{fig:Double} 
\end{figure}

From the structural perspective, the principal building blocks of ER are the fragments
of sheets and tubules made up
of double lipid bilayers \cite{Double}. We model the double bilayer as a self-avoiding
two-dimensional (2d) surface and the tubules as one-dimensional (1d) curves (see fig.~\ref{fig:Double}), in concurrence
with Ref.~\cite{Shemesh}. The edges of the double bilayer {\color{red}(red)} and the
tubules {\color{blue}(blue)} are decorated with reticulons~\cite{Reticulon}.
The quantitative calculation of the free energy for a self-avoiding fluid membrane with tubules
is complex: it involves the use theoretical models that have been motivated by experimental observations. 
We use a continuum-mechanics-based approach to study the phases, transitions, and morphologies
of such membranes under different biophysical conditions. 

\begin{figure}[!ht]
    \centering{
	\includegraphics[scale=0.25]{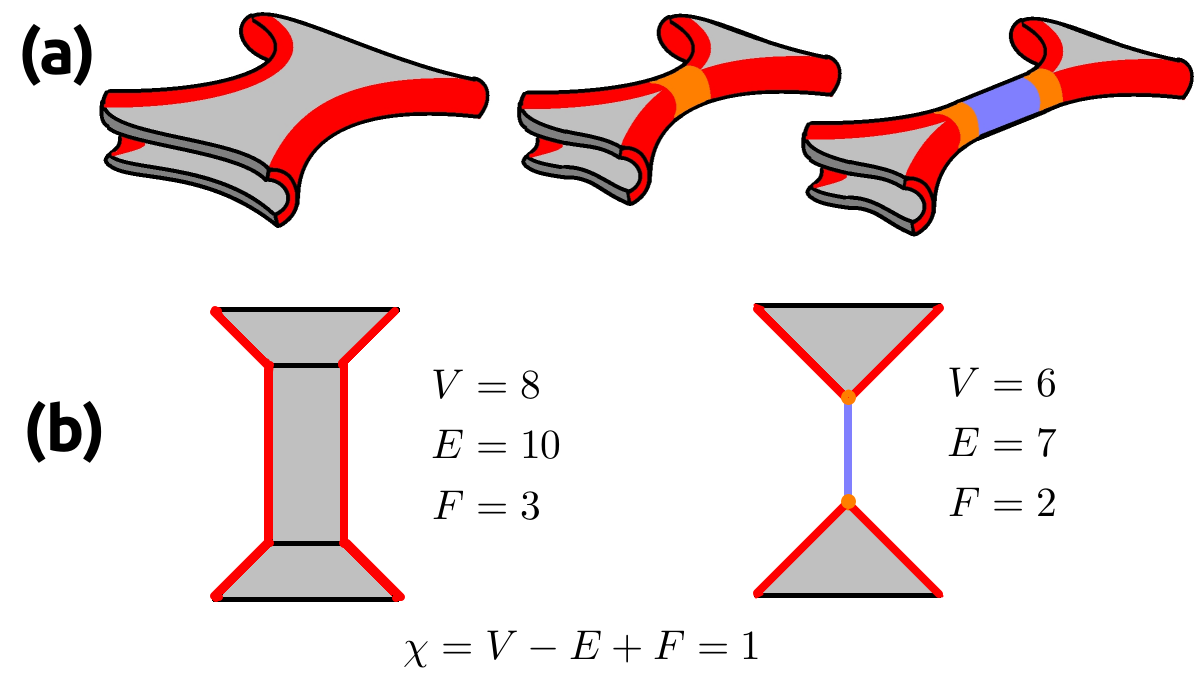}}
	\caption{Tubule formation: (a) A schematic where 
	         the edges of the double bilayer join together to form a tubule;
	         (b) The effective process when modelled with a 2d surface and 1d
              curve preserves the Euler characteristic $\chi$, with $V$-vertices,
              $E$-edges, and $F$-faces.}
	\label{fig:Top}
\end{figure}

In our study, we consider the Helfrich Hamiltonian~\cite{Helfrich} for a fluid
membrane together with line tension, and self-avoidance given below:
\begin{eqnarray}
    \mathcal{H} = 
        \underbrace{\gamma \oint_{\partial\mathcal{A}} ds \;\; + \;\;
	       \; \Gamma \int_\mathcal{T} ds}_{\text{Line-tension}} \;\; 
        + \;\; \underbrace{\iint_\mathcal{A} \left(  
	    \frac{\kappa}{2} \; H^2 + \kappa_G K \right) \sqrt{g} \;
            d^2\sigma}_{\text{Bending}}  
            \nonumber \\
        + \;\; \underbrace{ \mu \iint_\mathcal{A}  
	    \sqrt{g} \; d^2\sigma}_{\text{Surface tension}} \;\; 
        + \;\; \underbrace{4 \; \mathcal{B} 
	    \iint_\mathcal{A} \left( \left( \frac{\Delta}{\delta} \right)^{12}
            - \left( \frac{\Delta}{\delta} \right)^6  \right)
            \sqrt{g} \; d^2\sigma}_{\text{Self-avoidance}}  \,,
    \label{Eq:Model}
\end{eqnarray}
where $\mathcal{A}$ is the area of the surface elements and $\partial\mathcal{A}$ are
the corresponding boundaries; $\mathcal{T}$'s refer to the tubules [see
Fig.~\ref{fig:Double}]. We model the bending energy of the membrane 
by using the Helfrich terms~\cite{Helfrich}, 
which depend on the mean curvature $H$ and Gaussian curvature $K$. The surface- and line-tension 
energies arise from the stretching of the membrane's constituent molecules and
depend on the tension or stress applied to the membrane. The term with coefficient 
$\mu$ penalizes deviations from the membrane's preferred surface area, $\gamma$ is the
line tension of the surface edge, and $\Gamma$ is the line tension of the tubule.
The membrane's edges can coalesce to form tubules, as shown in Fig.~\ref{fig:Top}, so
a tubule is effectively two merged edges (see Fig.~\ref{fig:Top}), whence
$\Gamma = 2\gamma$. We neglect the energetics of the tubule-edge
junction, {\color{orange} orange}-region in Fig.~\ref{fig:Double}, because it 
involves higher-order gradients than those we retain in Eq.~\ref{Eq:Model}. 
The self-avoidance interactions between membranes are captured by
the Lennard-Jones-type potential~\cite{Lennard}, with coefficient $\mathcal{B}$, where
$\delta$ is the inter-layer spacing and $\left( 2^{1/6} \Delta \right)$ is the
preferred spacing.

For a free-standing membrane, with $\delta \gg \Delta$, a variation of the Hamiltonian~(\ref{Eq:Model}), with respect to its shape, leads to~\cite{Guven,JK}
\begin{eqnarray}
  4 \kappa H (H^2 - K) - 2 \mu H + \kappa \nabla^2 (2 H) = 0\,,
  \label{Eq:Bulk}
\end{eqnarray}
in the bulk; and the corresponding boundary conditions are:
\begin{eqnarray}
    \left.\left[ 2 \kappa H + \kappa_G \kappa_n \right]\right|_{\partial\mathcal{A}} &=& 0\,, \nonumber \\
    \left.\left[ -2 \kappa \partial_\perp H + \mu \kappa_n + \kappa_G \partial_\parallel \tau_g]
         \right]\right|_{\partial\mathcal{A}} &=& 0 \,, \nonumber \\
    \left.\left[2 \kappa H^2 + \kappa_G K + \mu + \gamma \kappa_g \right]\right|_{\partial\mathcal{A}} &=& 0\,,
    \label{Eq:BC}
\end{eqnarray}
where $\kappa_n$ and $\kappa_g$ are, respectively, the normal and the geodesic curvatures
of the boundary curve, with respect to (w.r.t.) the surface, $\tau_g$ is the torsion,
$\partial_\parallel \equiv \hat{\mathbf{t}} \cdot \nabla$ is the derivative
tangential to the boundary curve, and $\partial_\perp \equiv \hat{\mathbf{n}}_s \cdot \nabla$
is the derivative directed along the outward normal to the surface boundary.

Entropic contributions arise because of the membrane's ability to explore various conformations
via thermal fluctuations that lead to scale-dependent renormalizations of the coefficients in Eq.(~\ref{Eq:Model}). Hence,
we consider the free-energy contribution associated with the
fluctuations of the membrane. which can be written as
\begin{eqnarray}
    \mathcal{F} &=& - \ln{\mathcal{Z}} / \beta\,, \text{  where  } 
            \beta = 1/k_BT\,, \text{ and } \nonumber \\
    \mathcal{Z} &=& \int  \mathcal{D}[\mathcal{C}] \; \exp(-\beta \mathcal{H})\,, 
\end{eqnarray}
where $\mathcal{Z}$ is the functional integral over all possible
configurations of the membrane shape, with $\mathcal{D}[\mathcal{C}]$ 
the measure. We assume that the free energy has the same form as the
Hamiltonian~(\ref{Eq:Model}), with these renormalized coefficients. The inclusion of short-wavelength 
fluctuations renormalizes the bending moduli at the length scale $L$ as follows~\cite{David}:
\begin{eqnarray}
    \tilde{\kappa} &=& \kappa_0 - \frac{3}{4\pi \beta} \; \ln\left( L / a \right)\,,  \\
    \tilde{\kappa}_G &=& \kappa_{G0} + \frac{5}{6\pi \beta} \ln\left( L / a \right)\,,         
\end{eqnarray}
where $a$ is the molecular-length cut-off; $\kappa_0$ and $\kappa_{G0}$
are bare quantities, and $\xi_\kappa = a \; \exp\left( 4\pi \beta \kappa_0 / 3 \right)$ and 
$\xi_{\kappa_G} = a \; \exp\left( 6\pi \beta \kappa_{G0}/5 \right)$
as the corresponding thermal persistence lengths. The renormalized surface
tension is 
\begin{eqnarray}
    \tilde{\mu} = \mu_0 + \frac{\mu_0}{4\pi \beta \; \kappa_0} \ln\left(L/a\right) \,. 
\end{eqnarray}
The self-avoidance term also undergoes renormalization, with
\begin{eqnarray}
    \tilde{\Delta} = a + 1/\sqrt{\tilde{\kappa} |\tilde{\mu}| \beta^2}\,.
\end{eqnarray}
Furthermore, if $\delta = 2^{1/6} \delta_0 + \Delta$, then the quadratic expansion
of the compression term is $\sim  2^{17/3} \mathcal{B} \; (\Delta / \delta_0)^2 $.
Therefore,the renormalized layer compressibility is~\cite{Compression}
\begin{eqnarray}
    \tilde{\mathcal{B}} = \mathcal{B}_0 +  \frac{\pi^2 \tilde{\Delta}}{2^{2/3}
        \; \kappa \; \beta^2 \; (\tilde{\Delta}-\delta_0)^4}\,.
\end{eqnarray}

Explicit expressions for the renormalization of the line tension in fluid membranes
can be intricate and model-dependent, and there is no universally accepted formula for them.
So, for simplicity, we retain the bare value of this line tension.

In Sec.~\ref{sec:Results}, we consider different morphologies that we  
classify on the basis on their topologies. Furthermore, to simplify our analysis,
we study only periodic structures. Therefore, to compare the stabilities of the
different phases and their morphologies, we use the free energy per unit volume. All the
energy-minimizing solutions to the total free energy in Eq.~(\ref{Eq:Model})
are not amenable to analytic solutions because of the geometry-related
non-linearities. It is imperative, therefore, to look for approximate
numerical solutions. In the next Section, we give a brief
overview of the Surface Evolver package that we use for such numerical simulations.


\section{Methods: Surface Evolver}
\label{Sec:SE}

{\it Surface Evolver} (SE) is an open-source numerical package used in
computational physics and mathematics to model and simulate the behaviours
of surfaces and interfaces. Initially devised by Kenneth Brakke for mean-curvature-flow analysis~\cite{Brakke}, this tool has gained wide usage across diverse realms, prominently in exploring the physics of soap films, minimal
surfaces, and various interfacial phenomena.

Within SE, smooth 2D surfaces are represented as piece-wise-linear triangulated
surfaces embedded within Euclidean 3D space.

Key attributes of SE encompass:
\begin{enumerate}[label=(\alph*)]

    \item Numerical Surface Energy Minimization: Enables the discovery of minimal
    surfaces and configurations, while adhering to specified constraints.

    \item Versatile Geometry Simulation: Facilitates the simulation of intricate
    geometries, encompassing diverse surface types, shapes, and interactions.

    \item Dynamic Interface Simulation: Allows the study of surface evolution over
    time, yielding insights into growth, deformation, and phase transitions.

    \item Parameter Control: Offers manipulation of parameters like surface tension,
    boundary conditions, and constraints, thereby enabling the exploration of surface
    behaviours in varying environments and physical conditions.

\end{enumerate}

SE harnesses different numerical techniques [energy-minimization algorithms] to ascertain surface-equilibrium configurations. These algorithms calculate the most
energy-efficient shapes and structures for surfaces, with specified boundary
conditions and constraints. Its applications in material science include studies of
surface energies, stability assessments, and interface analyses. Moreover,
SE serves as a valuable tool in comprehending surface interactions within biological
systems, so it can aid in the modelling of biological membranes and cell structures.

\begin{figure}[!ht]
    \centering{
    \includegraphics[scale=0.7]{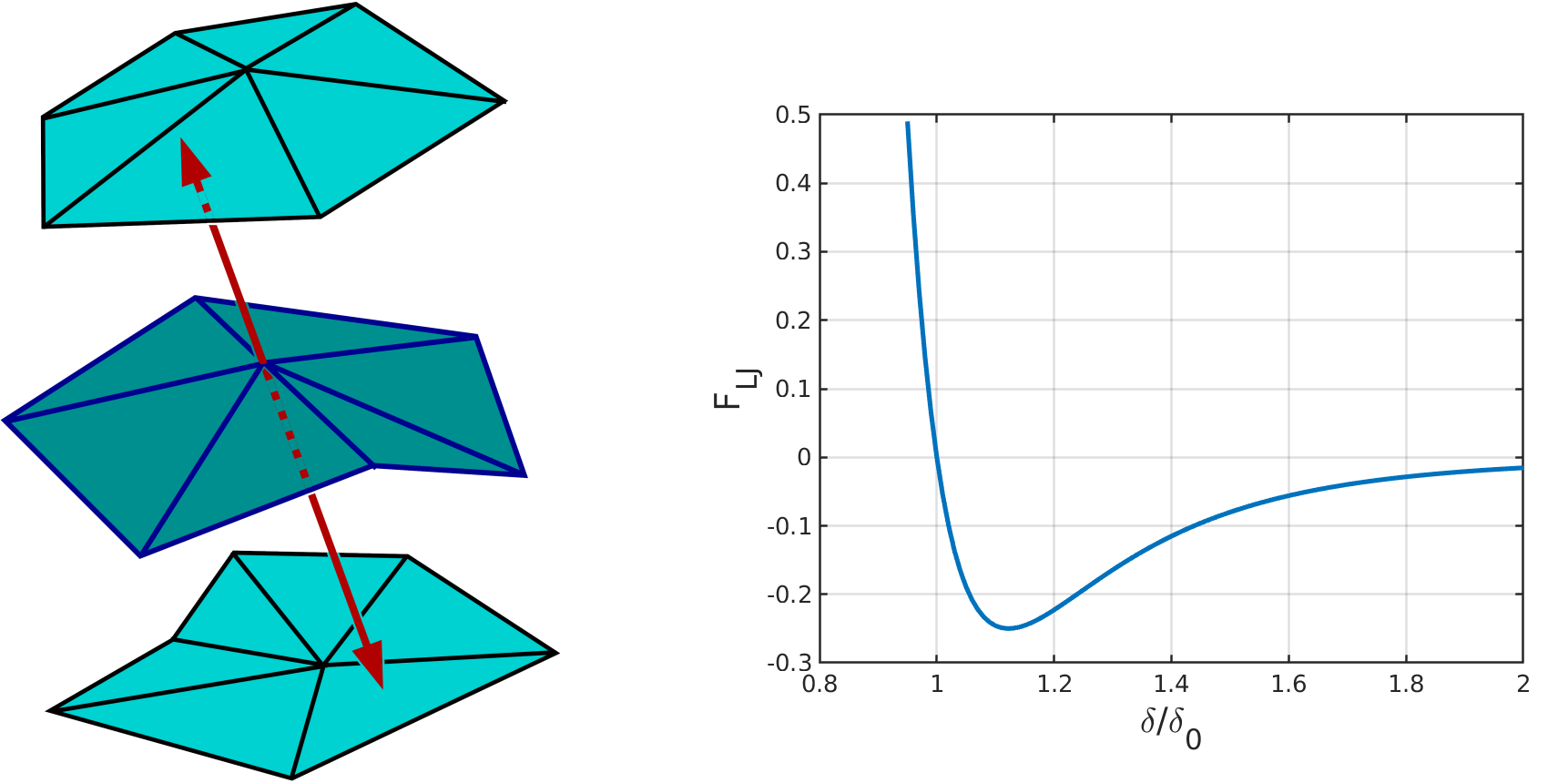}\\
    {(a)} \hspace{4cm} {(b)} }
    \caption{(a) Schematic diagram demonstrating the calculation of the distance between
    a vertex and neighbouring layers ($\delta$), along the vertex-normal.
    (b) Lennard-Jones-type potential between the surfaces. }
    \label{Fig:Compression}
\end{figure}

Working with SE typically involves scripting, in its dedicated programming
language, which allows users to define surface energies, constraints, and
boundary conditions. In our specific study, we introduce an auxiliary
term, represented as the self-avoidance $(\mathcal{B})$ in
Eq.~\ref{Eq:Model}, into the SE calculations to account for membrane
interactions. This term incorporates the interaction between membranes
by estimating the inter-layer distance, denoted as $\delta$, within the
lamellar phase. We determine this distance from triangles along the
vertex normal, as illustrated in Fig.~\ref{Fig:Compression}(a).

SE minimizes the energy by numerical shape variation, which is implemented by using
a conjugate-gradient descent~\cite{CGD}, w.r.t. the coordinate variables
of the vertices, as follows:
\begin{eqnarray} 
	\frac{\partial \mathbf{x}_\alpha}{\partial \tau}
	 = \mathbf{\Upsilon}_{\alpha\beta} \; . \; \frac{\delta
	\mathcal{F}}{\delta \mathbf{x}_\beta} \,, 
    \label{Eq:SE_Evolution}
\end{eqnarray} 
where $\mathbf{x}_\alpha$ is the 3d position vector of the vertex
labeled by $\alpha$, $\mathbf{\Upsilon}_{\alpha\beta}$
represents the vertex-dependent mobility matrix, and $\tau$ is the iteration
scale. Since we are modeling a fluid
membrane, the redundant evolution, leading to re-parameterization (in-plane
movement) of the vertices in the bulk, is eliminated by redefining the mobility
matrix as
\begin{eqnarray} 
	\mathbf{\Upsilon}_{\alpha\beta} \rightarrow
	\tilde{\mathbf{\Upsilon}}_{\alpha\beta} \; :=  \;\mathbf{n}_\alpha  \; . \;
	\mathbf{\Upsilon}_{\alpha\beta}  \,,
\end{eqnarray} 
where $\mathbf{n}_\alpha$ is the normal at the vertex labeled $\alpha$. The above
process enforces only the normal motion of surfaces. Furthermore, we set all
the vertex mobilities to be equal, {\it i.e.}, 
$\tilde{\mathbf{\Upsilon}}_{\alpha\beta} =  \Upsilon \; \delta_{\alpha\beta} 
\; \vec{\mathbf{n}}_\alpha$, where $\delta_{\alpha\beta}$ is a Kronecker
delta. The scale factor $\Upsilon$ is estimated by using Newton's method; and
it is proportional to the inverse of the Hessian~\cite{NLP}. We also note
that the Hessian of the edge vertices dominates over that of the bulk, leading
to slow the convergence of edge modes. To adjust for the offset, after
every iteration of the conjugate gradient descent we perform the variation
of only the edge vertices, while keeping all the vertices in the bulk fixed. 

\begin{figure}[!ht]
    \centering
    \includegraphics[scale=0.18]{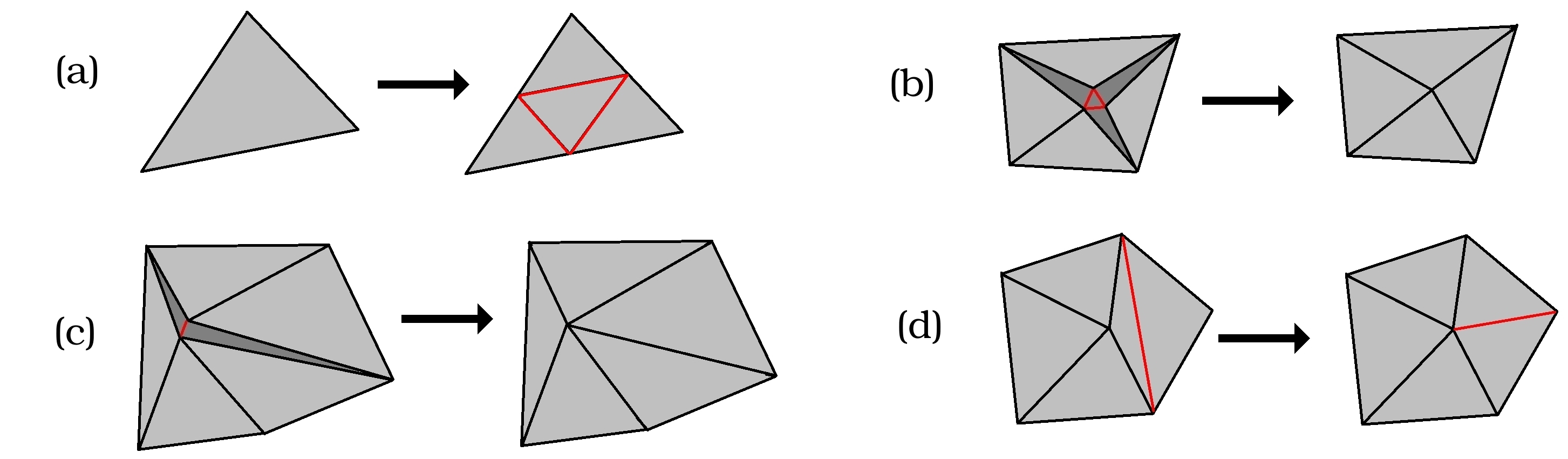}
    \caption{A schematic diagram of Pachner moves that we use to refine and re-triangulate the mesh:
    (a) refinement; (b) triangle weeding; (c) edge weeding; (d) edge flip.}
    \label{Fig:Pachner}
\end{figure}

We explore various morphologies detailed in Sec.~\ref{sec:Results}. These
distinct shapes possess unique topologies, which can be classified by their homology groups~\cite{Nakahara}. Our study
is initiated with triangular meshes that represent fundamental mesh structures
corresponding to each morphology. The process of energy minimization employs 
iterative triangle and edge refinements, alongside vertex evolution
guided by Eq.~(\ref{Eq:SE_Evolution}). This iterative refinement seeks stable surface
configurations, ensuring that vertex-coordinate changes remain within $5\%$ of
the unit cell size. Triangle refinements enhance surface resolution by
subdividing large triangles into smaller ones, effectively capturing
intricate surface details. We achieve triangle refinement
and reshaping by using Pachner moves~\cite{Pachner},
depicted in Fig.~\ref{Fig:Pachner}. These transformations preserve the
triangulation's topological properties. 

The systems that we consider are periodic in x, y, and z directions, with
orthorhombic or cubic unit cells; we implement this in SE by using the SYMMETRY 
option~\cite{Symmetry}. We obtain the energy
per unit volume to compare the stabilities of different phases of the system.


\section{Results}
\label{sec:Results}
    

Fluid membranes, which avoid self-intersections and incorporate edges,
possess the capacity to self-organize into diverse structural
configurations, to adapt to distinct environmental conditions, and
thus yield various phases [see, e.g., Refs.~\cite{Huse} and
~\cite{Menon}]. Our model introduces, in addition, the possibility of
tubule formation; this enriches the spectrum of possible phases in
the system. 

In our subsequent explorations of membrane stability, we consider cases in which $\kappa>0$
significantly surpasses $k_B T$. We focus on $\kappa_G = \kappa$; although the stability of fluid
membranes allow for both positive and negative values of $\kappa_G$,
we exxplore only the region $\kappa_G>0$.

We investigate the stability of diverse morphologies, by using our continuum model~(\ref{Eq:Model})
with renormalized coefficients (as discussed above). We utilize
Surface Evolver (SE) to capture the mean shapes. We focus on
the region $\mathcal{B} \sim \beta$; and $\delta_0$, which is 
equal to the double-layer width, induces a repulsive potential
between surfaces. This enables the fluctuations to counterbalance
attractive forces and allows surfaces to exist independently
without condensation.

Our exploration of the stabilities of different morphologies 
begins with the initialization of minimal triangular
meshes, with either orthorhombic or cubic symmetry, that are designed to
yield the desired topology. These meshes evolve
via SE to minimize the energy  [Eq.~(\ref{Eq:Model})]. As the
vertex-coordinate change falls below a $5\%$ threshold relative
to the unit-cell size, we iteratively refine and re-triangulate
the mesh by using the Pachner moves described above. We continue this until
further refinement fails to induce deformations [given the
stipulated threshold].

\begin{figure}[!ht]
    \centering{
	\includegraphics[scale=0.3]{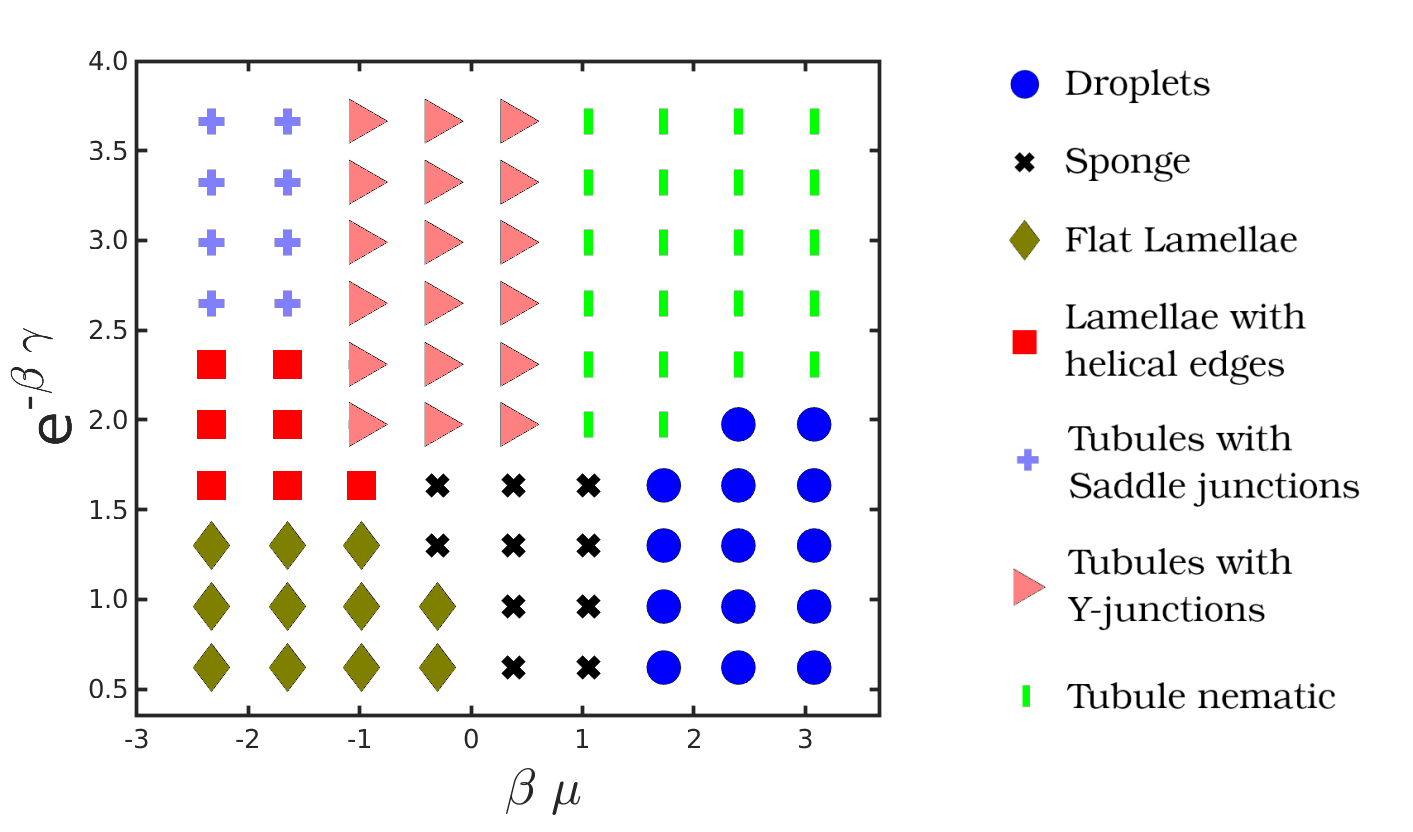} }
	\caption{The phase diagram for model~(\ref{Eq:Model}) with the representative value 
$\kappa = 5 \beta$ in the two-dimensional parameter space $[\exp{-(\beta \gamma)}, \beta \mu]$; this shows a variety of phases with non-trivial structures, that minimize the energy in different regions of this parameter space.
	         considered.}
	\label{fig:PD}
\end{figure}

We measure energies in units of $\beta$ and normalize
lengths by $\xi_\kappa$. Figure~\ref{fig:PD}
illustrates the phase diagram that we have obtained for model~(\ref{Eq:Model}) with the representative value 
$\kappa = 5 \beta$ in the two-dimensional parameter space $[\exp{-(\beta \gamma)}, \beta \mu]$; this shows a variety of phases with non-trivial structures, that minimize the energy in different regions of this parameter space. We proceed to present an in-depth analysis of these phases and the phase diagram; and we elucidate the interplay
between $\mu$ and $\gamma$ that stabilises different phases.

\begin{itemize}
    \setlength\itemsep{1.5em}

    \item[\ding{78}] {\it Lamellar phases:} 
    The system exhibits a strong inclination to form lamellae when
    $\mu$ has large negative values; this favours the and filling of the
    domain with layered surfaces (or lamellae). In describing the membrane stack
    within this phase, a persistence length ($\xi_\kappa$)
    signifies the average distance over which neighbouring membranes
    collide because of thermal fluctuations~\cite{CL}. These collisions
    result in a steric repulsion between the membranes and, consequently,
    renormalize the inter-layer spacing to 
    $\Delta \approx \xi_l \sim 1/ \sqrt{\kappa |\mu| \beta^2}$.

    \begin{itemize}
    \setlength\itemsep{1.5em}

    \item[\ding{111}] {\it Flat lamellae (FL):} When the energy ($\gamma$),
    associated with edges, becomes excessively high,
    the system actively avoids edges and the lamellae adopt a
    flat mean profile (see Fig.~\ref{Fig:FL}) to minimize the bending
    energy. In terms of topology, the phase comprises disconnected
    sheets, displaying a long-range orientational order, of
    the surface normal, and quasi-long-range
    positional order along this normal direction.

    If $\lambda_\perp$ and $\lambda_\parallel$ are the unit cell
    dimensions in the directions along the layer's normal and
    lateral directions, then the free-energy and the energy
    density reduce to:
    \begin{eqnarray}
        \mathcal{F}_{FL} &=& \tilde{\mu} \: \lambda_\parallel^2
            + 4 \: \tilde{\mathcal{B}}
            \left( \left(\frac{\tilde{\Delta}}{\lambda_\perp}
            \right)^{12} - \left(\frac{\tilde{\Delta}}
            {\lambda_\perp}\right)^6\right) \lambda_\parallel^2\,;
            \nonumber \\
            f_{FL} &=& \mathcal{F}_{FL} / ( \lambda_\perp 
            \lambda_\parallel^2 )\,;
    \end{eqnarray}
    $\lambda_\parallel=L=\xi_\kappa$; and we minimize
    $f_{FL}$ w.r.t. $\lambda_\perp$.
    \begin{figure}[!ht]
        \centering{
        \includegraphics[scale=0.3]{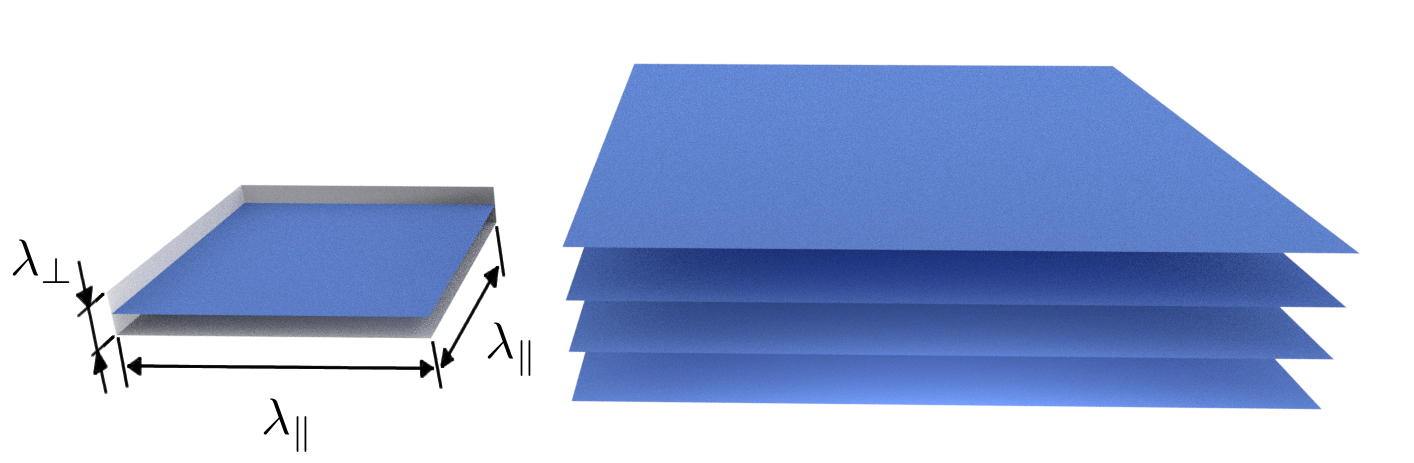}\\
        (a) \hspace{5cm} (b)}
        \caption{Structure of flat lamellae: (a) an orthorhombic
            unit cell; (b) the repetition of unit cells that form
            the final configuration of a flat lamellar phase.}
        \label{Fig:FL}
    \end{figure}

%

    \item[\ding{111}] {\it Lamellae with helical edges (LH):} 
    In the lamellar phase, as the $\gamma$ is reduced to a large
    negative value, the lamellae prefer to break and form sheets
    with edges. But simple circular holes or edges are
    unstable, so the system develops topologically non-trivial helical edges,
    exhibiting a morphology that resembles a staircase
    (see Fig.~\ref{Fig:Helicoid}(a)). This morphology consists of
    repeating, alternating layers, or terraces, connected via the
    helical edge. The helices are chiral; and opposite chirality
    helical edges can condense to form a square lattice as we show in
    Fig.~\ref{Fig:Helicoid}(b).

    We use the continuum elasticity model~(\ref{Eq:Model})
    to obtain the free energy. The lattice spacing, $\alpha$,
    of the helical edges is stabilized by the fluctuations; and
    it is determined self-consistently. Energetically, smaller values of
    $\alpha$ would lead to a higher density of edges and hence a lower
    energy. In our model, the lattice-spacing limit is set by
    the validity of the lamellar structure, which is 
    $\alpha\geq \xi_\kappa$, and we saturate this limit.
    Furthermore, by approximating the surface to be a helicoid,
    which is a minimal surface, we satisfy the force balance
    in the bulk [Eq.~(\ref{Eq:Bulk})]. The boundary
    conditions in Eq.~(\ref{Eq:BC}) relate the radius of the
    helical boundary to the interlayer spacing via~\cite{Yhat}
    \begin{eqnarray}
        - \frac{\tilde{\kappa}_G \; (\tilde{\xi}_l/2\pi)^2}
        {(r^2+(\tilde{\xi}_l/2\pi)^2)^2} + \frac{\tilde{\gamma}
        \; r}{(r^2+(\tilde{\xi}_l/2\pi)^2)} + \tilde{\mu} = 0\,,
    \end{eqnarray}
    where $r$ is the core radius (see 
    Fig.~\ref{Fig:Helicoid}(a)), and $\xi_l = 1/ \sqrt{\kappa
    |\mu| \beta^2}$. We begin with a coarse triangulated surface,
    as in Fig.~\ref{Fig:SE_Helicoid}(a), with $\lambda_\parallel
    = 2\alpha$; we then let SE evolve the vertices with multiple
    mesh refinements to reach the stable structure, shown in
    Fig.~\ref{Fig:SE_Helicoid}(c).
\begin{figure}[!ht]
    \centering{
    \includegraphics[scale=0.2]{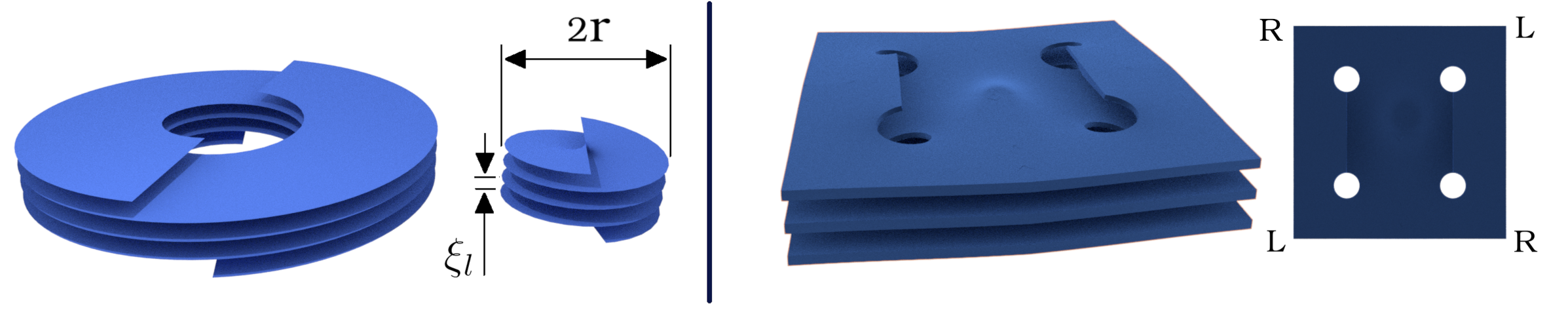}\\
    {(a)} \hspace{5cm} {(b)}}
    \caption{Schematic diagrams: (a) (right panel) the Helicoid core; 
    (left panel) the helicoid, with the core removed and showing the
    corresponding helical edge; (b) a unit cell of a pair of left(L)- and 
    right(R)-handed helicoids arranged on a square lattice
    [left panel: 3d view; right panel: top view].}
    \label{Fig:Helicoid}
\end{figure}

\begin{figure}[!ht]
    \centering{
    \includegraphics[scale=0.13]{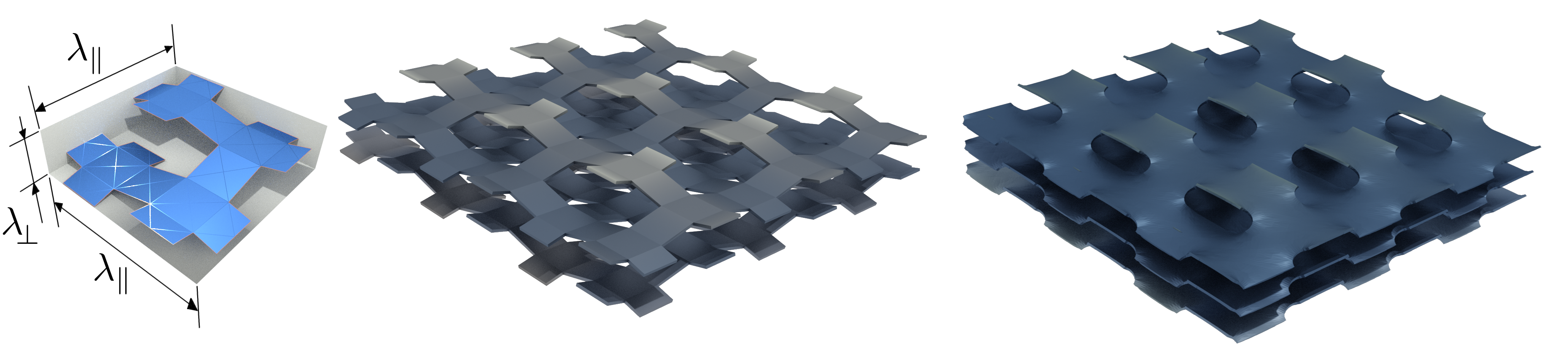} \\
    {(a)} \hspace{4cm} {(b)} \hspace{5cm} {(c)} }
    \caption{Schematic diagrams showing the structure of lamellae with helical edges: 
    (a) An orthorhombic unit cell with a triangulated surface
    that has two left- and two right-handed helical edges, as
    in Fig.~\ref{Fig:Helicoid}(b). This surface serves as the
    input initial condition for the SE. (b) The repetition of
    unit cells with orthorhombic symmetry. (c) The final
    configuration after energy minimization via SE.}
    \label{Fig:SE_Helicoid}
\end{figure}


\indent Studies such as those of Refs.~\cite{Terasaki} and \cite{Guven} associate the LH structures with the
rough ER, but they do not take into account the self-avoidance of the membranes. Instead,
to enforce a preferred inter-layer spacing
in the case of the staircase-like structure, they use a spontaneous geodesic-curvature term of the form $\int_{\partial \mathcal{A}} (\kappa_g - \bar{\kappa}_g)^2 \; 
dl$, where $\kappa_g$ and $\bar{\kappa}_g$ are the
geodesic curvatures of the boundary and the corresponding spontaneous value induced
by the protein that occupies the helical boundaries (and acts as springs with a preferred pitch).
But for a lattice of helicoids with lattice spacing greater than $\zeta_\kappa$,
thermal fluctuations render the structure unstable without self-avoidance.
Furthermore, the terms in
$\mathcal{F}_{geo}$ is of higher order compared to $\int K \; d^2\sigma$, as a
consequence of the Gauss-Bonnet theorem, which relates the
geodesic curvature and Gaussian curvature through the relation,
\begin{eqnarray}
	\int\int_\mathcal{A} \mathcal{K} \; \sqrt{g} \; d^2\sigma = 2\pi (1-g) 
	    + \oint_{\partial\mathcal{A}} k_g ds\,,
\end{eqnarray}
where $\mathcal{K}$ is the Gaussian curvature, $g$ is the genus of the surface,
and $k_g$ is the geodesic curvature. Now, if we use the $\oint k_g^2 ds$ term, we must
include, for consistency, higher-order terms in the Gaussian curvature term in the
bulk. Furthermore, the inclusion of the quadradic 
geodesic-curvature term stabilizes stacks of membranes with holes of radius
$\sim 1/\kappa_g$; these were not reported in Ref.~\cite{Terasaki}.


    \end{itemize}

    \item[\ding{78}] {\it Sponge phase:} 
    If $\kappa_G>0$, and we use a high value of $\gamma$ and a small value of $|\mu|$, the
    interlayer spacing decreases notably. Consequently, lamellae develop
    the propensity to merge, giving rise to intricately interconnected membrane
    networks that create a porous, labyrinthine structure called a sponge phase. In contrast to the
    flat lamellar phase, the sponge phase likes to bend and maximize negative
    Gaussian curvature. This phase adopts shapes akin to minimal surfaces with
    zero mean curvature without edges. The resulting sponge phase is characterized by a
    disordered and porous structure devoid of long-range order.

    We examine periodic
    sponge structures with cubic symmetry; these are often referred
    to as a \textit{plumber's nightmare} because of its complex
    morphology~\cite{Huse,Menon} (see Fig.~\ref{Fig:Sponge}(c)).
    To find the optimal plumber's-nightmare structure, we
    start with a triangular mesh, as shown in
    Fig.~\ref{Fig:Sponge}(a), and let the SE evolve the
    surface, iteratively refining the structure until
    the convergence criteria are satisfied. These
    structures exhibit three-dimensional crystalline
    order and are such that their area $\sim \lambda^2$, with
    a factor which we determine by using the plot in
    Fig.~\ref{Fig:SE_Sponge}, that we obtain by using SE. Since it is
    a minimal surface, $H=0$, and the integral of the
    Gaussian curvature per unit cell is independent of $\lambda$
    [it is a function only of the genus]. Furthermore,
    for the plumber's-nightmare structure, the entropic
    contribution is constant~\cite{Huse}. Hence, the total free
    energy density for the plumber's-nightmare structure for a 
    unit cell of size $\lambda(=L)$ is
    \begin{eqnarray}
        f_{SP} \approx A_1 \tilde{\mu} / \lambda + A_2 - A_3 
            \tilde{\kappa}_G / \lambda\,,
    \end{eqnarray}
    where $A_1$ is the slope of the linear fit in
    Fig.~\ref{Fig:SE_Sponge}(a), $A_2$ is the constant entropic
    contribution, and $A_3$ is the slope from
    Fig.~\ref{Fig:SE_Sponge}(b).
\begin{figure}[!ht]
        \centering{
        \includegraphics[scale=0.35]{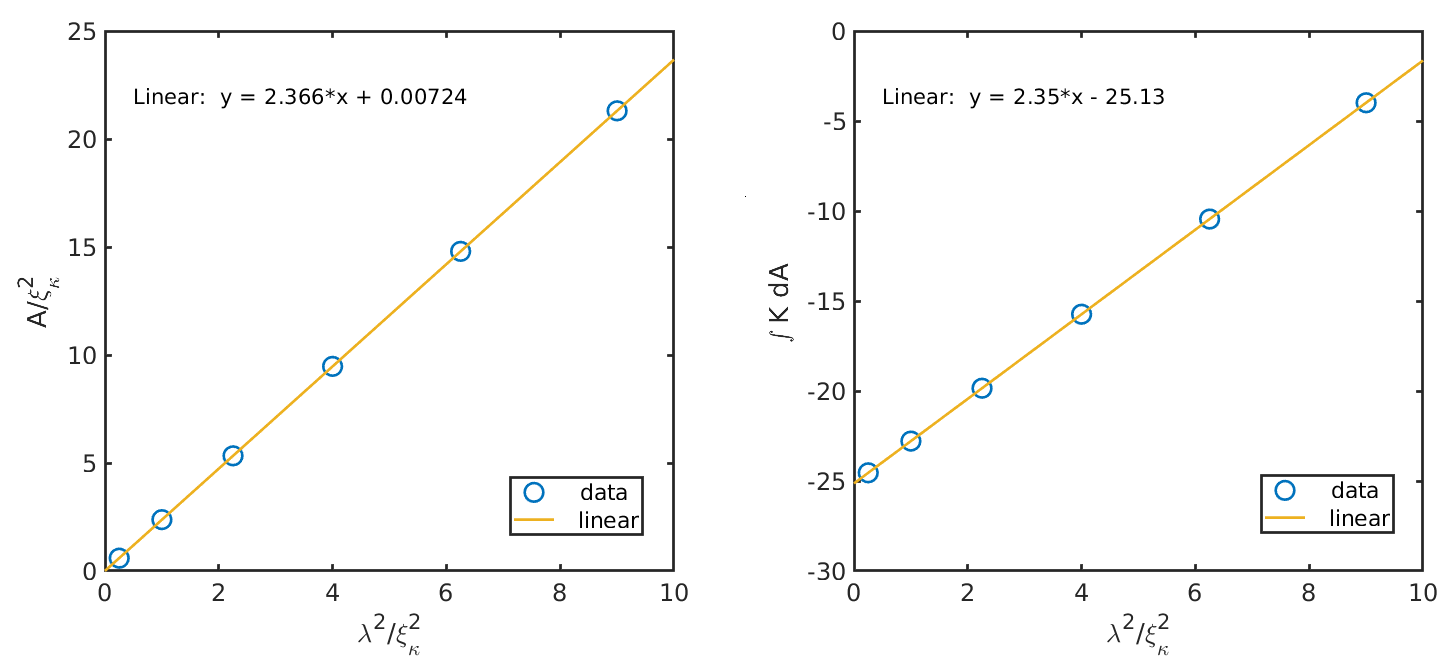}\\
        {(a)} \hspace{6cm} {(b)} }
        \caption{Scaling relations for the plumber's-nightmare phase
        obtained by using SE: (a) Plot of the area vs the unit-cell length
        ($\lambda$); (b) Plot of the total Gaussian curvature vs $\lambda$.}
        \label{Fig:SE_Sponge}
    \end{figure}

    Note that, for $\kappa_G>0$ and $\mu<0$, $f$ decreases
    with $\lambda$, so we have a UV runaway, and the stability of
    the structure is dictated by the short-distance
    cut-off; in our numerical study, the
    double bi-layer thickness provides this cut-off, i.e. 
    $\lambda \sim a$. 
    \begin{figure}[!ht]
        \centering{
        \includegraphics[scale=0.2]{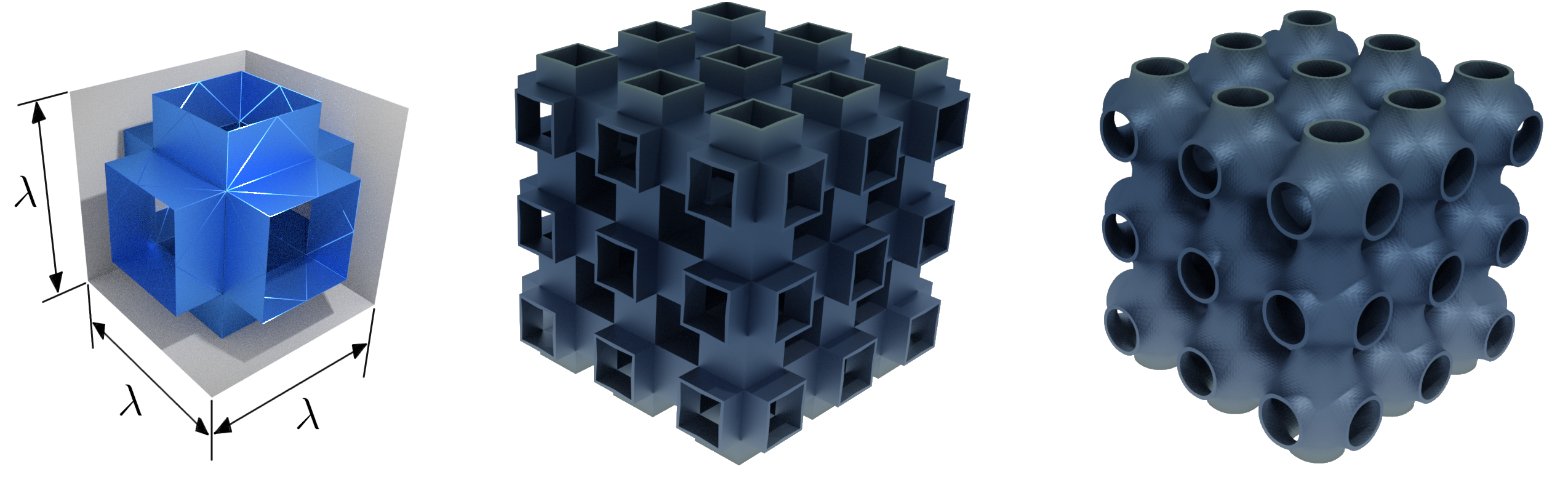} \\
        \hspace{0.5cm} {(a)} \hspace{4cm} {(b)} \hspace{5cm} {(c)} }
    \caption{Structure of the plumber's nightmare: 
    (a) Triangulated surface in a cubic unit cell that serves as the
    input to SE; (b) the repetition of unit cells with
    cubic symmetry; (c) final configuration after
    energy minimization via SE.}
        \label{Fig:Sponge}
    \end{figure}

    \item[\ding{78}] {\it Droplet phase:} 
    For positive $\mu$ and positive $\gamma$, the free energy
    can be reduced by the formation of {\it spherical droplets}
    whose radii are determined by fluctuations. The characteristic
    length scale, associated with $\mu>0$, determines the droplet
    size $\xi_\mu = 1/\sqrt{4 \pi \mu \beta}$; and the shape
    fluctuations determine the spacing between the droplets,
    which is of the order of $\xi_\kappa$. In general, for
    $\kappa_G>0$, the droplets proliferate [high
    genus topology]. For large $\mu$, the system prefers to
    avoid membranes by breaking up into dilute non-spherical
    droplets. We neglect the self-avoidance term because of
    the dilute-droplet assumption. To perform a self-consistent
    comparison between different morphologies, while keeping the
    analysis simple, we use a mono-disperse distribution of
    spheres with $\xi_\mu$ as the cut-off radius, and the
    lattice spacing of the order of $\lambda = \xi_\kappa$.
    Then, the corresponding free-energy density
    is~\cite{SafranDroplet}
    \begin{eqnarray}
        f_{D} \approx 4 \pi \: \tilde{\mu} \: \xi^2_\mu / \lambda^2
            + 4 \pi \: (\tilde{\kappa}/2 + \tilde{\kappa}_G )
            / \lambda^2 - \ln( \xi_\kappa / \xi_\mu ) 
            / (\beta \lambda^2)\,.
    \end{eqnarray}

    \item[\ding{78}] {\it Tubule phases:}
    When $\mu$ is positive and we have a negative value of $\gamma$, tubules dominate
    the morphology, which has string-like undulations and
    translational entropy of the junctions~\cite{Safran}.

    \begin{itemize}
        
        \item[\ding{71}] {\it Tubule nematic}:
        For large positive values of $\mu$, the system avoids
        forming surfaces and consists only of tubules.
        For the simple model that we consider, with only
        line tension for the tubules, they can align to
        give rise to nematic order, which optimizes packing.
        We consider a hexagonal packing of tubules and
        estimate the free-energy density.
                
        \item[\ding{71}] {\it Tubules with Y-junction}: 
        In the case of negative $\mu$ close to zero and negative
        values of $\gamma$, the tubules have string-like undulations and
        form 3-way Y-junctions, with a flat triangular patch
        of surface, as shown in Fig.~\ref{Fig:Y_junction}(a).
        The Y-junctions act as particles and give rise to translational entropy as in
        the Tlusty-Safran transition~\cite{Safran}. The break-up of
        the membrane into 3-way junctions is achieved through
        the transition in Fig.~\ref{fig:Top}. In our study, we
        consider the cubic arrangement of Y-junctions shown in
        Fig.~\ref{Fig:Y_junction}.

\begin{figure}[!ht]
    \centering{
    \includegraphics[scale=0.3]{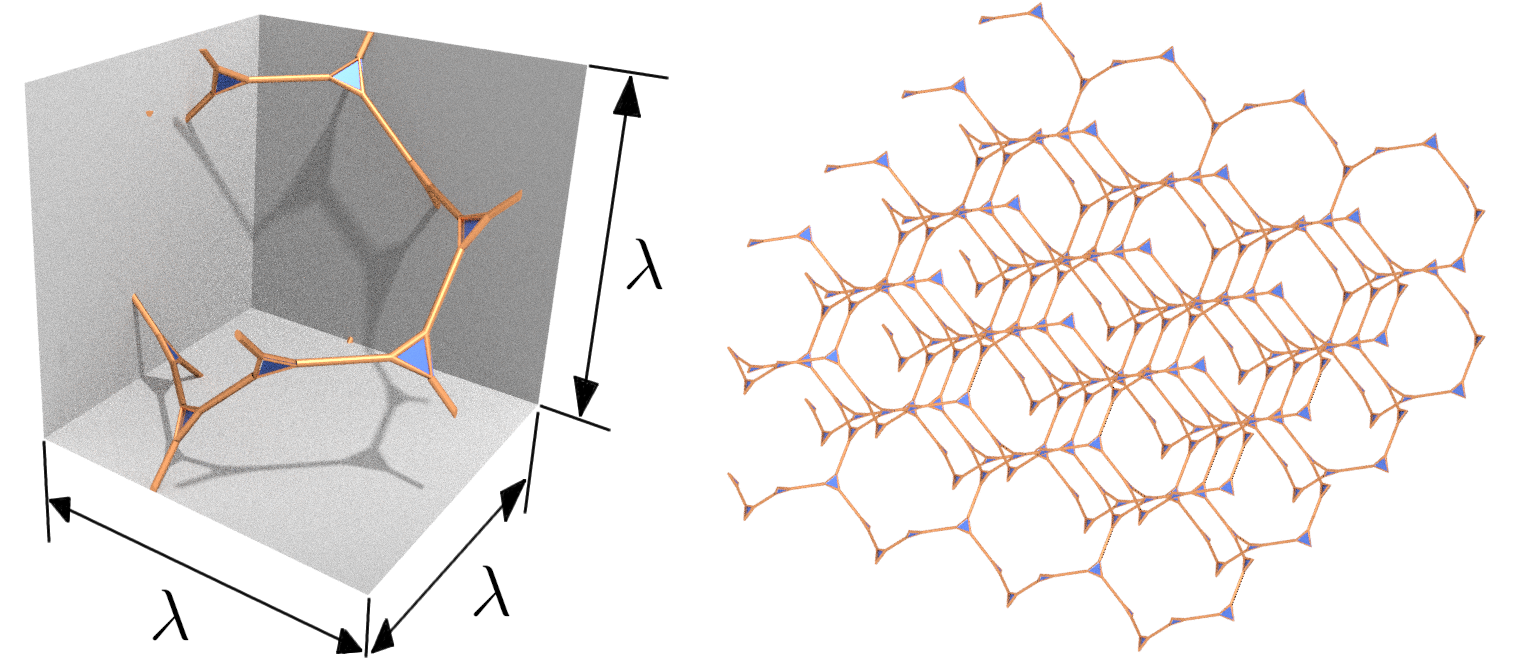} \\
    {(a)} \hspace{5cm} {(b)} }
    \caption{Schematic diagram showing the structure of tubules with Y-junctions: 
    (a) Triangulated surface in a cubic unit cell that serves as our
    input to SE. The tubules and surface edges are highlighted
    as 3d red tubes, but in SE they are 1d curves; (b) final
    configuration after energy minimization via SE, such that
    the unit cell is repeated with cubic symmetry.}
    \label{Fig:Y_junction}
\end{figure}

\begin{figure}[!ht]
    \centering{
	\includegraphics[scale=0.12]{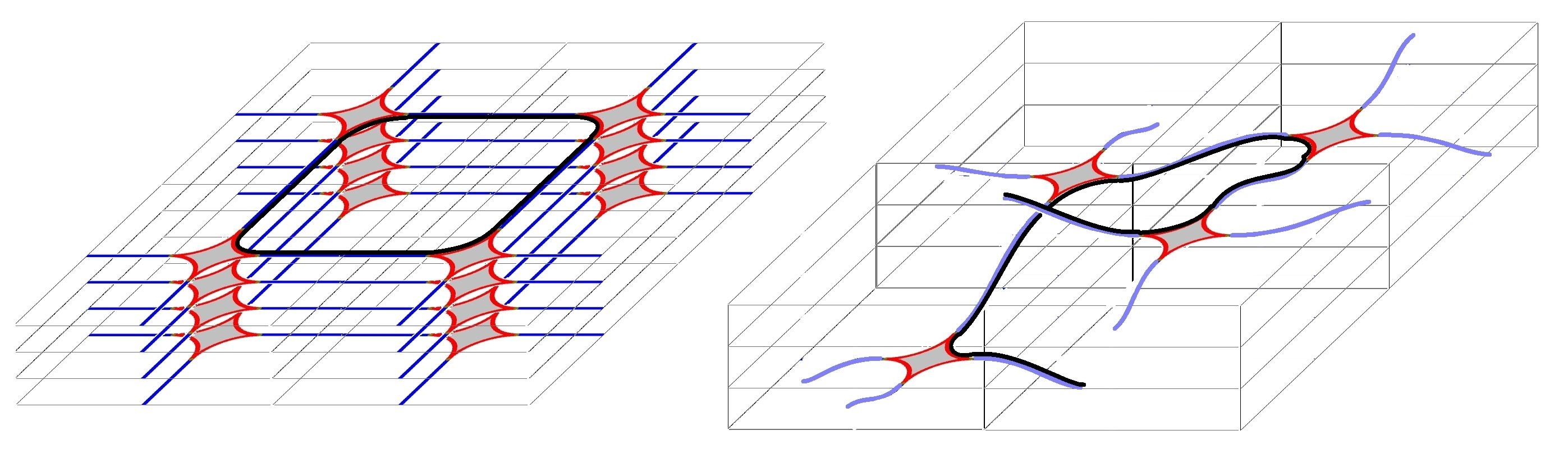}  \\
        {(a)} \hspace{4cm} {(b)} }
	\caption{A schematic diagram showing different possible topologies with 
            4-way junctions, with a Burger's loop highlighted
            in black: (a) fragments of flat membranes that
            connect to form a closed loop; (b) saddle
            membranes connected by tubules.}
	\label{fig:Fenestration}	    
\end{figure}
        
        \item[\ding{71}] {\it Tubules with saddle junctions}:
        If $\kappa_G>0$, for large negative $\gamma$ and
        small positive $\mu$, the edges in the staircase can grow and
        coalesce to form a network of tubules connected by small patches
        of sheets, to form {\it saddle-junctions}. The patches form
        pieces of saddle minimal surfaces that have negative Gaussian
        curvature and zero mean curvature (see Fig.~\ref{Fig:Saddle_SE}(c)).
        Here the shape and size of the patch are determined by the
        competition between the positive surface tension, $\mu$, which
        tries to minimize the area, and the negative contribution from
        the Gaussian curvature. To embed a negative Gaussian curvature
        surface in Euclidean three-dimensional space, the integral
        of the Gaussian curvature scales sub-linearly in the surface
        area~\cite{Nitsche} and hence leads to a finite size. We
        consider a lattice of such saddles and estimate their
        shape and size numerically.
        \begin{figure}[!ht]
            \centering{
            \includegraphics[scale=0.15]{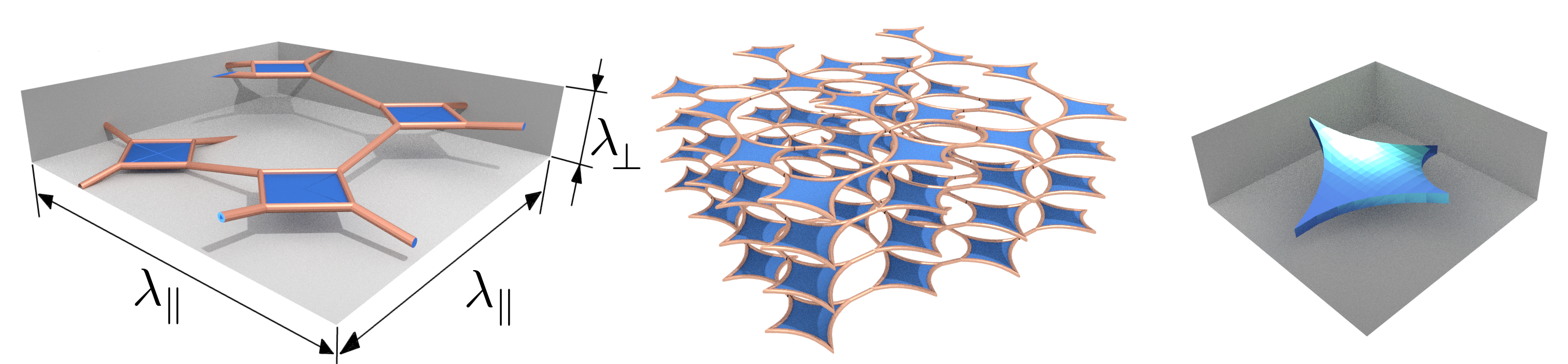}\\
            {(a)} \hspace{4cm} {(b)} \hspace{4cm} {(c)}}
            \caption{Schematic diagram showing the structures of saddle junctions: 
    (a) Triangulated surface in a unit cell that serves as our
    input to SE. The tubules and surface edges are highlighted
    as 3d red tubes, but in SE they are 1d curves; (b) final
    configuration after energy minimization using SE, such that
    the unit cell is repeated with orthorhombic symmetry;
    (c) the saddle shape of the junction with net negative
    Gaussian curvature minimizes the energy for $\kappa_G>0$.}
            \label{Fig:Saddle_SE}
        \end{figure}

        The Saddle junctions undergo a symmetry-breaking transition
        into the Y-junction phase, through the tubule formation
        process described in Fig.~\ref{fig:Top}, wherein a membrane
        breaks up into pieces of triangles as shown in
        Fig.~\ref{fig:Foam}. 
        \begin{figure}[!ht]
            \centering{
	       \includegraphics[scale=0.3]{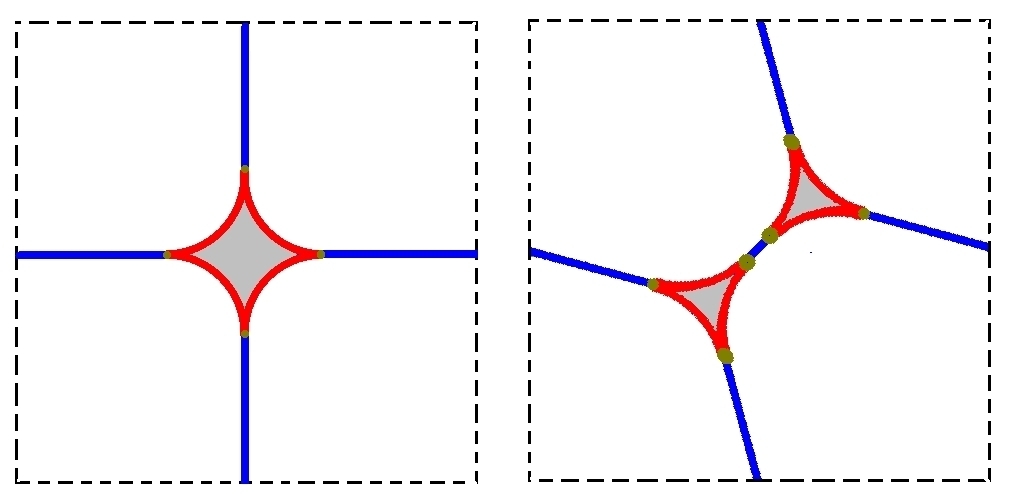} \\
               {(a)}  \hspace{3.5cm}  {(b)} }
	       \caption{Symmetry breaking transition from a saddle junction (a)
                to a Y-junction in (b).} 
	    \label{fig:Foam} 
        \end{figure}
        
    \end{itemize}

\end{itemize}


The phase diagram in Fig.~\ref{fig:PD} summarizes the analysis presented above.
As observed in Ref.~\cite{Terasaki}, the lamellar phase with helical edges
corresponds to the rough ER and the tubule phases correspond to the
smooth ER. Through our model, we highlight that there exists a structural
transition between the lamellae with helical edge, saddle-junctions, and
Y-junctions such that the interior of the double bilayer, the lumen,
maintains contiguous connectivity.



%

\section{Conclusions}
\label{sec:Conclusion}

Our study demonstrates that the diverse structural phases witnessed within the
endoplasmic reticulum (ER) can be effectively replicated by using our model~(\ref{Eq:Model}) that encompasses
interacting fluid membranes featuring edges. By employing energy-minimization techniques
and variational methods, we derive the stable shapes characterizing these distinct
structures.

The resultant phase diagram, derived through numerical optimization, unveils transitions
from lamellar configurations to helicoidal stacks, closely resembling the experimentally
observed structures within the rough endoplasmic reticulum (RER). Furthermore, our
findings highlight the existence of homotopy-preserving transitions, illustrating
transformations between helicoidal stacks, saddle junctions, and Y-junctions.

\subsection{Discussion}

The intricate nature of endoplasmic reticulum (ER) structures, coupled with the diverse
array of factors influencing its morphology poses a significant challenge in
constructing a comprehensive phase diagram that shows the varied ER morphologies
under diverse conditions. In this context, several critical points require attention:
\begin{enumerate}[label=(\alph*)]
    \item The diverse morphological variations across different cell types and
        physiological contexts.
    \item Understanding the interconversion and remodelling dynamics between
        tubules and sheets.
    \item Investigating the influence of cellular activity on ER morphology.
    \item Unraveling the intricate interplay among lipid bilayers, proteins,
        and cellular signalling pathways.
    \item Deciphering the role of proteins in sculpting and determining ER
        morphology.
\end{enumerate}

Advancements in super-resolution microscopy, live-cell imaging techniques, and computational
modelling continuously contribute valuable insights into ER morphology. Recent research
endeavours have made strides in elucidating phase transitions and morphological alterations
within the ER. However, the construction of a detailed phase diagram that comprehensively
integrates these findings remains a work in progress.

Limitations of our model: The ER system is, in general, active and far from equilibrium. 
 While our work identifies key influential conditions and
factors impacting ER morphology, crafting an exact phase diagram that integrates dynamic
activity and out-of-equilibrium components remain an ongoing challenge.
The inclusion of active non-equilibrium phenomena can be accomplished by using
coloured-noise-based membrane fluctuations, as discussed, e.g., in Ref.~\cite{Sarasij};
such studies have not yet yielded the structures we have obtained.

\section{Acknowledgements}

YH and JKA express gratitude to V.A.~Raghunathan for valuable inputs and insightful
discussions. JKA thanks Chaitra Hegde for engaging in relevant discussions.

We acknowledge the Department of Science and Technology, New Delhi, for their support
through the DSTO1359 grant. RP acknowledges the support received from the Science and
Engineering Research Board (SERB) and the National Supercomputing Mission (NSM). We
also extend our gratitude to SERC (IISc) for providing computational resources.

We acknowledge the use of AI tools like Grammarly and ChatGPT for sentence rephrasing
and grammar checks. Subsequently, the material underwent meticulous proofreading to
ensure precision and rectify any errors. Our process includes thorough reviews and
edits to ensure accuracy, relevance, and coherence in the finalized text.



\end{document}